\documentclass[12pt]{iopart}

\usepackage{iopams}
\usepackage[T1]{fontenc}
\usepackage[utf8]{inputenc}
\usepackage{lmodern}
\usepackage{graphicx}
\usepackage{dcolumn}
\usepackage{bm}
\usepackage{breqn}
\usepackage{setspace}
\usepackage[footnotesize]{caption}
\usepackage{xcolor}
\usepackage[normalem]{ulem}
\usepackage{supertabular}
\begin{document}

\title[Substructure energies and probabilities in 2D-silica]{Relating local structures, energies, and occurrence probabilities in a two-dimensional silica network}

\author{Projesh Kumar Roy$^{1,2}$ \& Andreas Heuer$^{2,3,*}$}

\address{$^1$ NRW Graduate School of Chemistry, Wilhelm-Klemm-Stra{\ss}e 10, 48149 M\"unster, Germany}
\address{$^2$ Institute f\"ur Physikalische Chemie, Westf\"alische-Wilhelms-Universitat M\"unster, Corrensstra{\ss}e 28/30, 48149 M\"unster, Germany}
\address{$^3$ Center for Multiscale Theory and Computation, Westf\"alische-Wilhelms-Universitat M\"unster, Corrensstra{\ss}e 40, 48149 M\"unster, Germany}

\ead{andheuer@uni-muenster.de}
\vspace{10pt}
\begin{indented}
\item[]\today
\end{indented}


\begin{abstract}

Recently, it became possible to experimentally generate and characterize a very thin silica system on a substrate which can be basically described as a 2D random network. The key structural properties, in particular related to the ring statistics, could be numerically reproduced by performing molecular dynamics simulations with an appropriately chosen 2D force field. Using a maximum entropy formulation it is shown that the probability distribution of the individual rings and triplets can be related to the ring and triplet energies, respectively, obtained from the simulations. Using additional Lagrange parameters, the correct average properties of random networks are guaranteed. In agreement with previous work, based on distributions of complementary rings and triplets, respectively, one finds a Boltzmann-type relation albeit with an effective temperature which largely deviates from the bath temperature. Furthermore, it is shown that the ring and triplet energies can be estimated based on the properties of their average inner angles. This calculation supports, on a quantitative level, the previously suggested angle mismatch theory. It suggests that correlations among adjacent rings originate from the net mismatch in the inner ring angles in a triplet of rings. By taking into account an average effect from the surrounding rings of a triplet, an even better estimate of the correlations can be provided. That approach is also applied to estimate the Aboav-Wearie parameter.

\end{abstract}


\submitto{\JPCM}
\maketitle


\section{Introduction}
\label{sect:Introduction}

A random network is a collection of elements of different size and shapes which are connected randomly in a closed fashion. Although termed as `random', it is still possible to distinguish  different random networks by observing average topological properties. The standard deviation of the elementary sizes, the Aboav-Wearie parameter \cite{AboavMetallography1970, WearieContempPhys1984}, the topological correlation parameter \cite{DelannayJPhysA1992}, the Lewis constant \cite{LewisAnatRec1928}, or the Von Neumann constant \cite{NeumannMetalInterf1952} are a few important parameters that can define typical statistical properties of a random network. Several empirical rules were proposed for calculating these average properties \cite{AboavMetallography1970, WearieContempPhys1984,LewisAnatRec1928,RivierPhilMagB1985,LemaitrePhilMagB1993}.

In such periodic two-dimensional random networks, the average ring size is six \cite{WearieContempPhys1984}, i.e.,

\begin{equation}
\sum_n nP(n) = 6.
 \label{eqn:avringsize}
\end{equation}

Here $P(n)$ is the probability that a ring of size $n$ is observed. This important constraint is called the `Euler-Poincar\'e  characteristics'. A key aspect of the random network statistics deals with the correlations among adjacent ring sizes. As the total number of vertices is constant due to equation \ref{eqn:avringsize}, many relations can be derived between the ring size and the ring sizes of the neighbors \cite{PeshkinPRL1991, RivierDisorder1993}. An important example is a sum rule, relating the average size of the neighbors ($m(n)$) of a ring of size $n$ and the variance ($\sigma_r^2$) of the ring size distribution in 2D, namely the `Wearie sum rule' \cite{WearieMetallography1974}.

\begin{equation}
 \sum_n nm(n)P(n) = 36 + \sigma_r^2
 \label{eqn:wearie_sum_rule}
\end{equation}

To determine the correlations in the immediate neighborhood of a ring, the Aboav-Wearie parameter ($a_{AW}$)\cite{AboavMetallography1983, ChiuMatCharac1995} is an important quantity. It is an empirical measure of such correlations which comes from Aboav's rule relating the size of a ring ($n$) and the average size of the neighbors of that ring ($m(n)$),

\begin{equation}
 m(n) = 6 - a_{AW} + \frac{6a_{AW} + \sigma_r^2}{n}
 \label{eqn:Aboav}
\end{equation}

For a positive value of $a_{AW}$, this equation suggests that the rings with small sizes prefer rings with large sizes in their neighborhood and vice versa.
However, these descriptions of correlations do not reflect the microscopic origin of possible inter-ring correlation effects. It was suggested that the inter-ring correlations are dependent on the contact angles of the rings \cite{AboavMetallography1983, WearieContempPhys1984, RivierPhilMagB1985} in a network. It was also shown that the internal angles of the rings can have important effects on the dynamical aspects of the network, e.g. the area-growth of soap-bubbles \cite{NeumannMetalInterf1952, GlazierPRL1989, WeariePhilMagB1991}.

\begin{figure*}[!htb]
\centering
  \includegraphics[width=8cm, height=7cm]{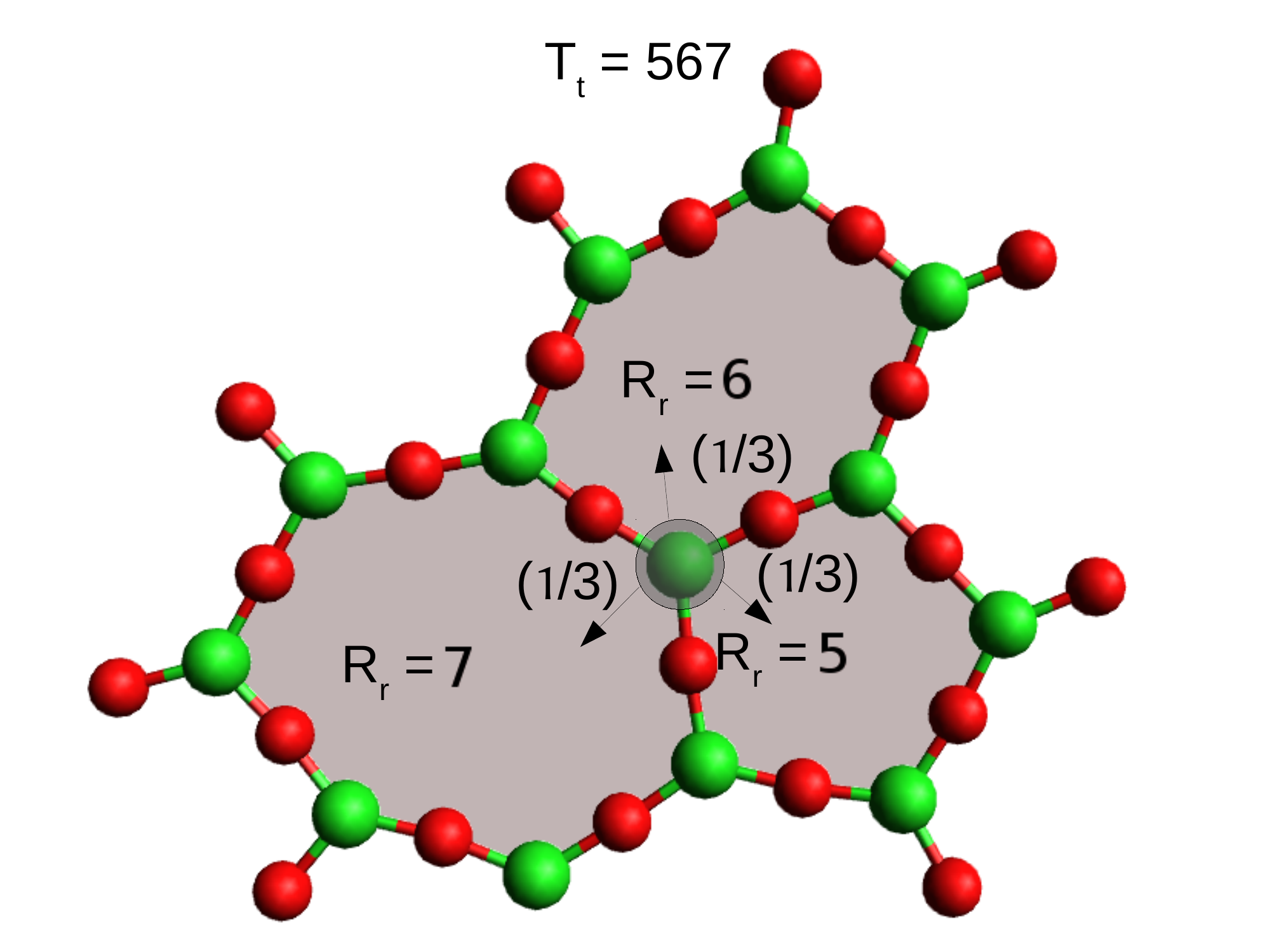}
  \caption{Sample of a triplet of rings in 2D silica, sharing a common corner. Image partially reproduced from  \cite{RoyPCCP2018} by permission of the PCCP Owner Societies. As discussed below in sect. \ref{subsect:One_third_rule_ring} the effective energy of each Si particle is equally distributed to the three rings, connected with this particle.  The triplet energy is calculated by adding all energies of each ring. The ring size is denoted as $R_r$, the triplet size as $T_t$.}
  \label{fig:triplet}
\end{figure*}

In 2011, Lichtenstein et al. were able to resolve a two-dimensional random network composed of silica on a Ru(0001) metal support by using scanning tunneling microscopy (STM)\cite{HeydeAngew2012}. Similar results have been obtained by Huang et al. on a graphene support by using scanning transmission electron microscopy (STEM)\cite{HuangNanoLett2012}. A local part is sketched in Fig. \ref{fig:triplet}. In particular, the statistics of individual rings and triplets could be obtained. Naturally, the silica rings fill the complete system in 2D plane. Thus, the so-called two-dimensional silica (or 2D-silica in short)is an excellent example of an atomistic random network of the most common glass former present in nature. Despite being 3D, the layered structure of this allotrope gives rise to an effectively two-dimensional material due to symmetry relations between the layers. For this system, either the silicon or the oxygen coordinates can be measured in a two-dimensional plane via STM \cite{HeydeJPhysChemC2012}.

So far, the theory of random networks was mostly developed using topological parameters \cite{RivierDisorder1993, RivierPhilMagB1985, RivierJPhysA1982} in macroscopic systems. The physics of macroscopic networks, e.g., soap bubbles or living cells, are quite different from that of atomistic systems. At first glance, both networks may look equally random. However, the network dynamics in soap bubbles is controlled by the gas diffusion across boundaries which results in phenomena like coarsening or Ostwald ripening \cite{RivierDisorder1993}. A steady state in a soap bubble network can be described as a minimum surface energy network. On the other hand, ring networks in atomistic systems like silica are controlled by the energy loss due to bond length and/or angle deviations.

Previously, a graphene type model was proposed to analyze the atomistic random networks \cite{WilsonJPhysCondMatt2012,  WilsonPhysRevB2013, WilsonJPhysCondMatt2014, VinkJChemPhys2014}, which is also relevant for two-dimensional silica. Recently, it was shown that for such a model, the correlations in between the rings can be described by calculating the dispersion of a `topological charge' parameter of a ring with the geometric distances in between the rings \cite{MahdiPhysRevE2016}. In this study, it was found that the 6-rings are less correlated with its surroundings than other ring-sizes.

To evaluate the thermodynamic aspects of the correlation effects, one needs an expression that relates the inter-ring correlations to the system energy. However, we still lack a strict energetic description in the atomistic random network theories. In case of a bulk silica ring network, Rino et al. \cite{VashishtaPhysRevB1993} had devised a `harmonic' potential approximation to formulate a ring-energy parameter, which predicts the ring size probabilities quite well. These parameters depend on the energy penalties from the deviations of inner ring-angles and deviations of bond-lengths of the edges from the average values. However, these energy parameters are not derived in a self-consistent way to capture at the same time the total energy of the system. In case of zeolites, a more force-field based approach was taken by Sastre et al. \cite{SastreJPhysChemB2006} to derive the ring-energy parameter by calculating the total interactions of all silicon and oxygen atoms present in a ring. However, one still needs to consider effects of other atoms situated at larger distances. Also, in this definition of ring-energy, the sum of the energies of all rings will not be equal to the total energy of the system.  B\"uchner et al. \cite{HeydeJnonCrysSolid2016} suggested an `angle-mismatch' approach. Here, the mismatch of the total inner ring-angles from $360^\circ$ at any vertex is directly related to the likelihood of finding this vertex and thus to inter-ring correlations. Indeed, a good agreement with experimental data was observed.

Recently, an atomistic force field in 2D has been developed which allows one to reproduce many structural features, seen experimentally \cite{RoyPCCP2018}, on a quantitative level. On this basis, we aim to find thermodynamic access to the properties of random networks.  A first step was already taken in \cite{RoyPRL2019}. There it was shown that it is possible to define, e.g., ring energies based on the atomistic simulations and to relate these energies to their occurrence in equilibrium simulations. The key result is that the resulting Boltzmann distribution contains an effective temperature which, for rings, is significantly smaller than the actual equilibrium temperature. On a qualitative level, this was related to the correlations among adjacent rings and the general observation that correlation effects may give rise to effective temperatures even in equilibrium situations. However, this approach was only able to predict the properties of pairs of rings with an average size of six. Thus, it was not possible to predict the occurrence of 5-rings and 7-rings separately but only the product of their probabilities. Similarly, the occurrence of `triplets', was predicted. A triplet denotes the three rings, which are connected to a single silicon atom (see figure \ref{fig:triplet}).

The scope of this paper is twofold. First, we introduce a theoretical approach to predict the occurrence of all ring sizes separately by taking into account the constraint of the average ring size and test this approach for the actual numerical data. Following previous work on networks, we will choose a maximum entropy approach and introduce an appropriate Lagrange parameter to take care of the correct average ring size. This approach is also generalized to the prediction of the occurrence of individual triplets. Second, following and extending the previous work \cite{RoyPRL2019} to estimate the local stress, relevant for rings and triplets, we formulate explicit expressions to predict ring and triplet energies with a minimum number of adjustable parameters. These predictions can then be directly compared with the actual ring and triplet energies, respectively.

The structure of the paper is as follows. In section \ref{subsect:Maximum_entropy_background} we present a brief overview of the maximum entropy method. In section \ref{sect:Simulation} we describe our model system and the simulation methods used to sample various equilibrium ring-networks at various temperatures. Also, our approach to define the energies of rings and triplets is outlined. Next, in section \ref{sect:Ring_energy_theory} we formulate the theory of ring and triplet statistics with the help of maximum entropy methods. Based on this approach, in section \ref{sect:Ring_energy_results} we analyze the simulation results. In section \ref{sect:Angle_mismatch_theory}, we show how the ring and triplet energies can be connected to the average of the inner angles of the rings and triplets, respectively. We end with a discussion and an outlook.

\section{Maximum entropy methods in random networks}
\label{subsect:Maximum_entropy_background}

In a random network, the average topological properties from large samples are usually the only available pieces of information. A closer theoretical understanding of these systems has been obtained via `maximum entropy' methods, as initially suggested by Jaynes \cite{JaynesPhysRev1957} following Shannon's interpretation of entropy \cite{ShannonBellSysTechJ1948} $S = -\sum P_ilnP_i$ as a measure of `uncertainty' present in the system. Maximizing the total uncertainty under a set of restrictions imposed on the system (externally or internally), one can derive the theoretical expectation values of any variable using a statistical mechanical approach, similar to equilibrium statistical thermodynamics. The probability distribution of such systems can be derived by using multiple Lagrange parameters which control the average values of the corresponding observables. Of course, for this method to be valid, one has to affirm a `statistical equilibrium' where micro-reversibility and ergodicity are maintained in the ensemble \cite{RivierPhilMagB1985}.

In physical random networks, such as living cells or soap bubbles, the restrictions are generally related to the topology of the system. Rivier et al. \cite{RivierJPhysA1982} used maximum entropy methods with only topological restrictions and proposed a more fundamental formulation of the statistics of random networks. A detailed description of these restrictions can be found in  \cite{RivierPhilMagB1985}, which together with maximum entropy methods yields different rules such as  Lewis's area rule \cite{RivierJPhysA1982,RivierPhilMagB1985}, the perimeter rule \cite{RivierPhilMagB1985}, or Aboav's rule \cite{PeshkinPRL1991, DelannayJPhysA1992}.

\section{Simulation approach}
\label{sect:Simulation}

\subsection{A model system : Two dimensional silica}
\label{subsect:2D_silica}

For the case of 2D-silica we have parametrized \cite{RoyPCCP2018} a 2D {two-body} Yukawa type force-field \cite{AlejandreJChemPhys2005, AlejandreCondMattPhys2012, AlejandreJChemPhys2012} 

\begin{equation}
  V_{ij}(r_{ij}) = \left [\left ( \frac{\sigma_{ij}}{r_{ij}} \right )^{12} + \left ( \frac{q_{ij}}{r_{ij}} \right ) \exp(-\kappa r_{ij}) \right ].
  \label{eqn:yukawa}
\end{equation}

With this force-field the different pair-correlation functions, the ring and triplet statistics as well as angle distributions such as for the $\angle SiSiSi$ angle distribution are in close agreement to the experimental data.  For our present study of ring-distributions the latter observable is of descriptor of the ring-size.

We mention in passing, that also for bulk-silica a successful reproduction of many structural and dynamical properties can be performed with just a two-body force-field \cite{Beest90,HeuerPhysRevLett2004}, albeit also force-fields with three-body terms have been employed; see \cite{Paramore08}  for a review.

The details of the simulation procedure can be found in  \cite{RoyPCCP2018}. In brief, the simulation involves 32 `Si' and 48 `O' particles in a squared simulation box of length 19.67 {\AA} with periodic boundary conditions. We frequently minimize and filter all defect states where at least one Si particle does not have 3 O particles as neighbors or at least one O particle does not have 2 Si particles within a given cut-off distance. The resulting defect-free configuration contains 16 rings and an average ring size of 6 in all frames. We simulated the system for a range of temperature (0.014 $\leq$ T $\leq$ 0.018). For the details of these dimensionless units to actual temperatures, we again refer to  \cite{RoyPCCP2018}.

\subsection{Energy of a ring and triplet}
\label{subsect:One_third_rule_ring}

Recently \cite{RoyPRL2019}, we have introduced a definition of ring and triplet energies without any prior assumption about bond-lengths or angle dependencies. In short, this was achieved in two stages. First, the atomistic energies of the O particles were allocated in equal parts to the two connected Si particles. This way an `effective energy' of the Si particles was given as

\begin{equation}
\epsilon_{\textsf{\scriptsize  Si,eff}} = \frac{1}{2} \left [ \epsilon_{\textsf{\scriptsize Si}} + \frac{1}{2}\sum_{i = 1}^3 \epsilon_{\textsf{\scriptsize O}}^i \right ].
\label{eqn:sieff}
\end{equation}

Then, the energy $\epsilon_{\textsf{\scriptsize Si,eff}}$ was equally divided in three parts and each part contributes to the energy of one of the three connected rings (see figure \ref{fig:triplet}). The total energy of a ring, $\epsilon_r$, was thus the sum of all the contributions made from the attached Si particles. It follows that the sum of all ring energies corresponds to the total energy $E_{\textsf{\scriptsize tot}}$ of that configuration. For this purpose, it was essential to introduce the prefactor of (1/2) in equation (\ref{eqn:sieff}) to account for the fact that all pair energies are counted twice.

\begin{center}
 \small
 \captionof{table}{Average values of the ring-energies for different ring-sizes($R_r$) at T = 0.015. The standard deviation $SE_r$ for the estimated average energy per particle $E_r/R_r$  is calculated as ($\sigma_r\sqrt{2<\tau_r> /N_r}$), where $\sigma_r$ is the standard deviation of the normalized energy distribution, $N_r$ is the total count of $R_r$-sided rings in our simulations, and  $<\tau_r>$ the average lifetime of a ring, obtained from reference \cite{RoyPCCP2018}. This factor takes care of the reduced information due to correlations of subsequent configurations}.
 \label{tbl:Ring_av_energy}
 \begin{tabular}{|c|c|c|c|}
  \hline
  Ring size  & Probabilities & Av. Energy &   standard deviation \\
  ($R_r$)     & ($P_r$)          & ($E_r/R_r$)           & ($SE_r $) \\
  \hline
  4 & 0.051 & -0.27891 &    0.000034 \\
  \hline
  5 & 0.273 & -0.28038 &    0.000023 \\
  \hline
  6 & 0.381 & -0.28053 &    0.000021 \\
  \hline
  7 & 0.226 & -0.28032 &    0.000027 \\
  \hline
  8 & 0.059 & -0.27994 &   0.000038 \\
   \hline
  9 & 0.011 & -0.27948 &   0.000061 \\
  \hline
\end{tabular}
\end{center}

\begin{figure*}[!htb]
\centering
  \includegraphics[width=8cm, height=8cm]{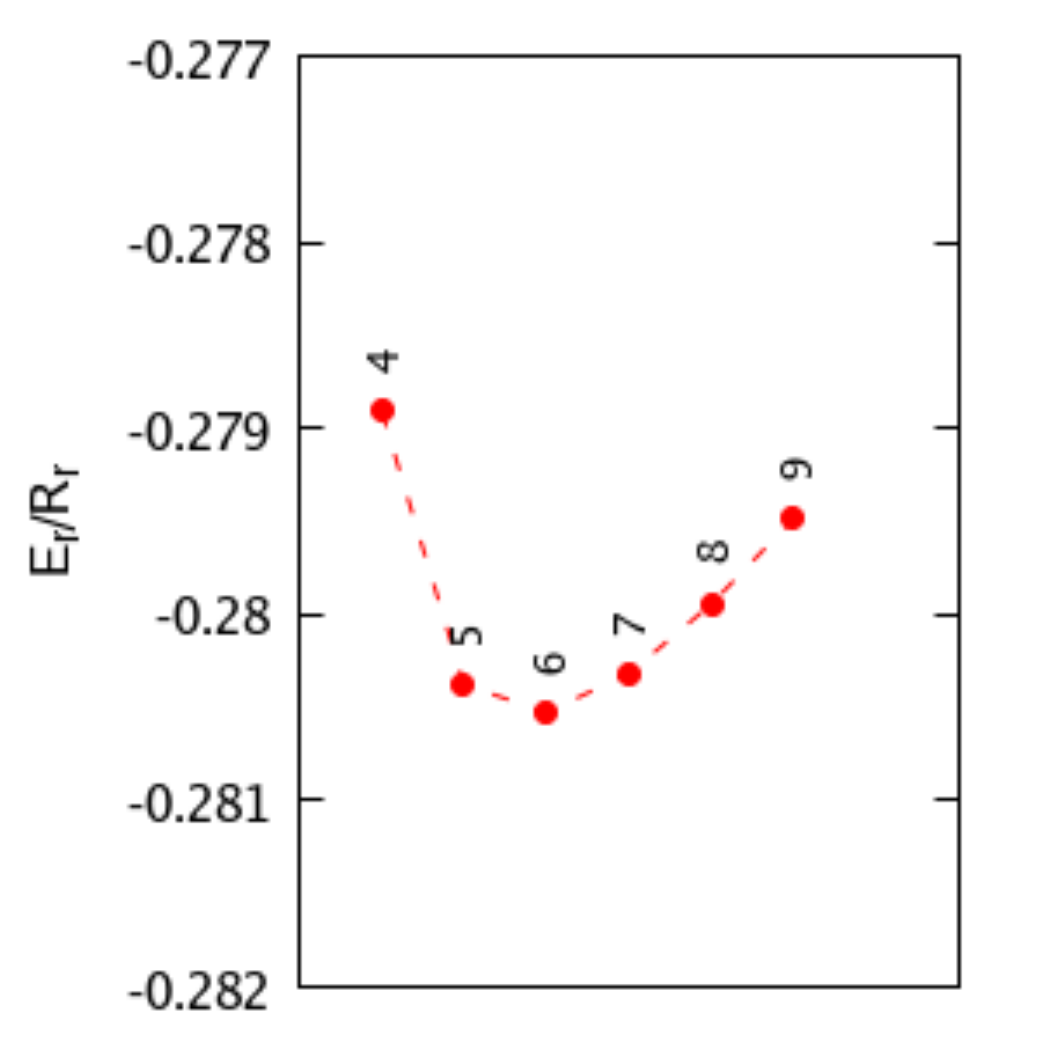}
  \caption{Average ring-energy ($E_r$) relative to the ring size ($R_r$) at T = 0.015. }
  \label{fig:ring_av_energy}
\end{figure*}

\begin{figure*}[!htb]
\centering
  \includegraphics[width=16cm, height=6cm]{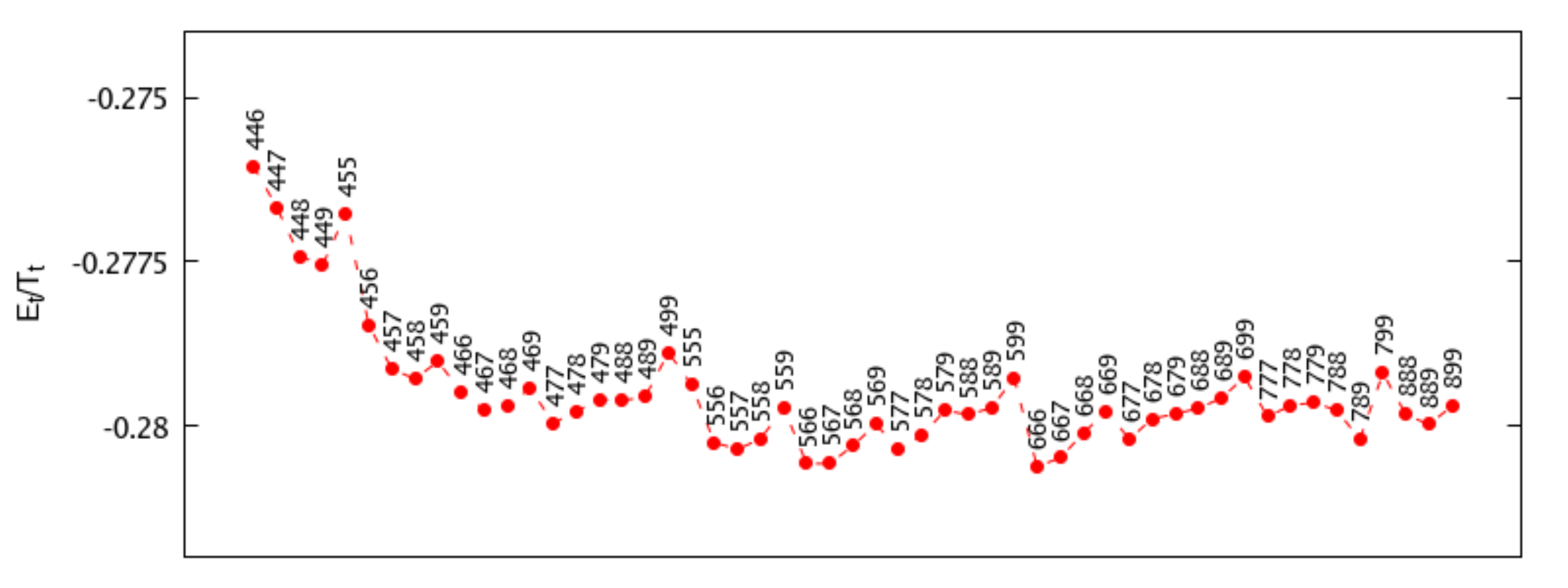}
  \caption{Average triplet-energy ($E_t$) relative to the triplet size ($T_t$) at T = 0.015. }
  \label{fig:triplet_av_energy}
\end{figure*}

The average energy of a ring-size $R_r$ from all states is expressed as $E_r$. To define the energy of triplet of size $T_t$ (discussed later), we simply add the energies of all constituting rings. This way, the central Si particle has a maximum contribution for the triplet energy. The average energy of a triplet of size $T_t$ from all states is expressed as $E_t$. The results for $E_r/R_r$ and $E_t/T_t$ are shown in figure \ref{fig:ring_av_energy} and figure \ref{fig:triplet_av_energy}, respectively. Although the variation of $E_r/R_r$ is small for the different ring sizes relative to the absolute values, it is exactly this variation which gives rise to, e.g., the dominance of 6-rings. 

For both ring (see table \ref{tbl:Ring_av_energy})  and triplet energies, the standard error of the mean was found to be significantly smaller than the energy differences, normalized per particle. Hence, we can state that the average energies are very well-defined. Also, as shown in reference \cite{RoyPCCP2018}, the finite size effect can be excluded as the distribution of ring sizes is the same within statistical noise for a 80 particle system and a 500 particle system. This helps us to calculate the ring/triplet-energies, extracted from defect-free configurations, with sufficient statistical accuracy \cite{RoyPCCP2018}.

It should be noted that there is more than one way to redistribute the atomistic energies to a ring while keeping the total energy unchanged. In the current energy redistribution procedure, we not only incorporate the energies of all Si and O particles placed on the boundary of the ring, but also the O particles which are connected to the ring-Si particles but not the ring itself. These O particles should have an indirect effect on the ring energy, as their energies are strongly correlated with the connected Si particles. The rational behind our choice is that we regard the Si particles as the ring-defining objects and thus, in the first step, remove the explicit contribution of O particles. 

\subsection{Comparison with the DFT energies}

We now investigate how closely the energy parameters of rings can be compared with the energies optimized by DFT. Previously, Lichtenstein et al. \cite{HeydeAngew2012} studied the average energy requirement for introducing a 5-7 ring pair in a network optimized via DFT. In a fully crystalline structure, pairs of 5-7 rings were introduced via arbitrary bond-rotations. The average increase in the total energy for introducing a 5-7 ring pair was about 0.17 eV. We re-estimated this energy requirement with the average ring energies, calculated at $T = 0.015$, observed in the simulations. The results are listed in table \ref{tbl:DFT_comparison}.

\begin{center}
 \small
 \captionof{table}{Comparison of the estimated energies with the DFT data from figure 4(c) in  \cite{HeydeAngew2012}. We show the energy difference from the structure 1 in this reference. The DFT data was supplied in kJ/mol, which is here expressed in eV.}
 \label{tbl:DFT_comparison}
 \begin{tabular}{|c|c|c|c|}
  \hline
  Rings added &  DFT energy diff. & Est. energy    \\
  (in bracket)&  from Str. 1      & diff. from rings   \\
              & (eV)              & (red. units)    \\

  \hline
  +(5,7)           &  0.18 & 0.0022  \\
  \hline
  +2$\times$(5,7)  &  0.34 & 0.0044  \\
  \hline
 \end{tabular}
\end{center}

The comparison suggests that the energy scale is roughly 77 eV. This value is larger than the previous rough estimate of 56 eV in  Ref.\cite{RoyPCCP2018}. Although the scale is of no relevance in the present work, one may keep the value, determined in this work, as the more realistic one. It is because, we determine the scale from the difference in energies, not the absolute energies where systematic errors cannot be ruled out.

\section{Theory of ring energy distribution}
\label{sect:Ring_energy_theory}

\subsection{Statistics of rings with different sizes}

Going beyond our previous work \cite{RoyPRL2019}, we show how to formulate the individual ring statistics. For this purpose, we define $N_r$ as the total count of ring-size $R_r$, found in the simulation. Then the constraint of total energy can be written as

\begin{equation}
\sum_r N_r E_r =  E_{\textsf{\scriptsize tot}}
\label{eqn:constraint1}
\end{equation}

Note that the use of the usual Boltzmann distribution $P_r \propto \exp(-\beta E_r)$ would largely overestimate the presence of large rings since to first approximation $E_r$ is proportional to the ring size.

This problem can be solved if we use equation (\ref{eqn:avringsize}) as a second constraint, giving rise to an additional Lagrange parameter. For better comparability with the first term in the exponential, we aim to express this constraint in terms of energies. For this purpose we introduce $E_{\textsf{\scriptsize av}}$ as the average energy per particle, i.e. $\langle \epsilon_{ \textsf{\scriptsize Si,eff}}\rangle$. Then we define

\begin{equation}
E_r^0 = \frac{R_r E_{\textsf{\scriptsize av}}}{3}
\label{eqn:define_Er0}
\end{equation}

as the average energy per ring, if the energy per particle would not depend on the ring size. The factor 3 appears because for each particle the energy contributes to three different rings. Now the constraint equation (\ref{eqn:avringsize}) can be rewritten as

\begin{equation}
\sum_r N_r E_r^0 = (1/3) \sum_r N_r R_r E_{\textsf{\scriptsize av}} = 2 E_{\textsf{\scriptsize av}} \sum_r N_r  = E_{\textsf{\scriptsize tot}}
\label{eqn:constraint2}
\end{equation}

The second equality expresses that the average ring size is 6. In the final relation, we have explored that the total number of rings is just half the number of Si particles for an average ring size of 6. Note that each particle is connected to 3 rings. Since a ring has on average 6 particles, the factor reads 3/6=1/2, which cancels with the factor of 2 in equation (\ref{eqn:constraint2}).

Finally (see also below) the first constraint equation (\ref{eqn:constraint1}) is rewritten (together with equation (\ref{eqn:constraint2})) as

\begin{equation}
 \sum_{r} N_r [ E_r - E_r^0 ] = 0
 \label{eqn:ring_energy_correlation}
\end{equation}

Based on both constraints the estimation for the probability $P_r$ is given by

\begin{equation}
 P_r \propto  e^{-\beta^{\scriptsize \texttt{eff}}_{\scriptsize \texttt{ring}} [E_r - E_r^0]
                 -\eta_{\scriptsize \texttt{ring}} E_r^0 }
 \label{eqn:ring_new_boltzmann}
\end{equation}

By using the expression $\beta^{\scriptsize \texttt{eff}}$ as the Lagrange parameter, reflecting an inverse effective temperature, we anticipate a key result, reported in \cite{RoyPRL2019}. There is has been shown that one obtains a Boltzmann relation, albeit with a non-standard inverse temperature due to thermodynamic effects on small scales. 

Both Lagrange parameters have a very different meaning. The first parameter captures the impact of average local stresses as reflected by the ring energy $E_r$ relative to the average energy of all particles. The second constraint directly guarantees an average ring size of six. Obviously, the definition of both Lagrange parameters is not unique, e.g. one might have written $ P_r \propto  \exp[-\beta^{\scriptsize \texttt{eff}}_{\scriptsize \texttt{ring}} E_r -\tilde{\eta}_{\scriptsize \texttt{ring}} E_r^0 ]$. It turns out, however, that exactly with the above choice the value of $\eta_{\scriptsize \texttt{ring}}$ turns out to be temperature independent (see below).

Formally, the statistics of the elementary building blocks, i.e., the rings, can be regarded as a grand-canonical ensemble where the chemical potential (related to the terms proportional to $E_r^0$) guarantees an average ring size of six. Since a ring with ring size $R_r$ is an elementary system,  we do not have to take into account an additional factor to represent the density of states.

\subsection{Generalization to triplets}

In analogy to the ring energies we define $E_t$ as the average energy of all triplets $t \in {i,j,k}$, detected in simulations. The total number of triplets in a configuration is the same as the total number of corners. The size of a triplet is defined by the sum of its ring sizes, $T_t = R_i + R_j + R_k$. If no correlations are present, the probability of simultaneously finding three rings in the network is a simple multiplication of the individual ring probabilities and the average triplet energy is a simple sum of the average ring energy parameters as

\begin{eqnarray}
 P_t^{predicted} = f_t  P_i P_j P_k \nonumber \\
 E_t^{predicted} = E_i + E_j + E_k
\label{eqn:Triplet_predicted}
 \end{eqnarray}

Here, $i \leq j \leq k$ for the triplet notation. $f_t$ is the permutation factor given by

\begin{eqnarray}
  f_t & = 1 \condition{for $i=j=k$} \nonumber \\
      & = 3 \condition{for $i=j\neq k$} \nonumber \\
      & = 6 \condition{for $i \neq j \neq k$}
 \label{eqn:pf}
\end{eqnarray}

However, these predictions do not take into account that the connection of the three rings by a single point gives rise to energetically more or also less favorable combinations of rings \cite{HeydeJnonCrysSolid2016, RoyPCCP2018}. Furthermore, since several particles of a triplet belong to more than one ring one expects energetic correlations which invalidate the use of the estimation of $E_t$ as suggested in equation (\ref{eqn:Triplet_predicted}) (see below).

Similarly to  equation (\ref{eqn:avringsize}), triplets also fulfill a topological restriction, expressed as

\begin{equation}
 \sum_{t} P_t T_t = \frac{1}{2}(36 + \sigma_r^2).
 \label{eqn:const_average_trip_size}
\end{equation}

Equation (\ref{eqn:const_average_trip_size}) is basically identical to  equation (\ref{eqn:wearie_sum_rule}) which was already used as a constraint related to energy in previous work \cite{RivierGarching1998, RivierDisorder1993}. Note that there is a key difference between equation (\ref{eqn:avringsize}) and  equation (\ref{eqn:const_average_trip_size}). From knowledge of the distribution of triplets, it is possible to derive the distribution of ring sizes. Thus, the right side is not a constant but also depends on the distribution of triplets. If e.g., only 666 triplets are present, one would naturally have $\sigma_r^2 = 0$. As a consequence, one may expect that the prediction of probability distributions works better for rings than for triplets.

Similar to the derivation of  equation (\ref{eqn:ring_new_boltzmann}), we can express the topological constraints by an energetic constraint and end up with the modified Boltzmann probability distribution

\begin{equation}
 P_t \propto f_t e^{-\beta_{\scriptsize \texttt{triplet}}^{\scriptsize \texttt{eff}} [ E_t - E_t^0 ] - \eta_{\scriptsize \texttt{triplet}} E_t^0}
 \label{eqn:trip_new_boltzmann}
\end{equation}

In analogy to $E_r^0$, the energy $E_t^0$ denotes the energy of that triplet under the assumption that all particles possess the same energy.

\section{Prediction of ring and triplet probabilities}
\label{sect:Ring_energy_results}

\subsection{Statistics of rings and triplets}

\begin{figure*}[!htb]
\centering
 \includegraphics[height=6cm,width=12cm]{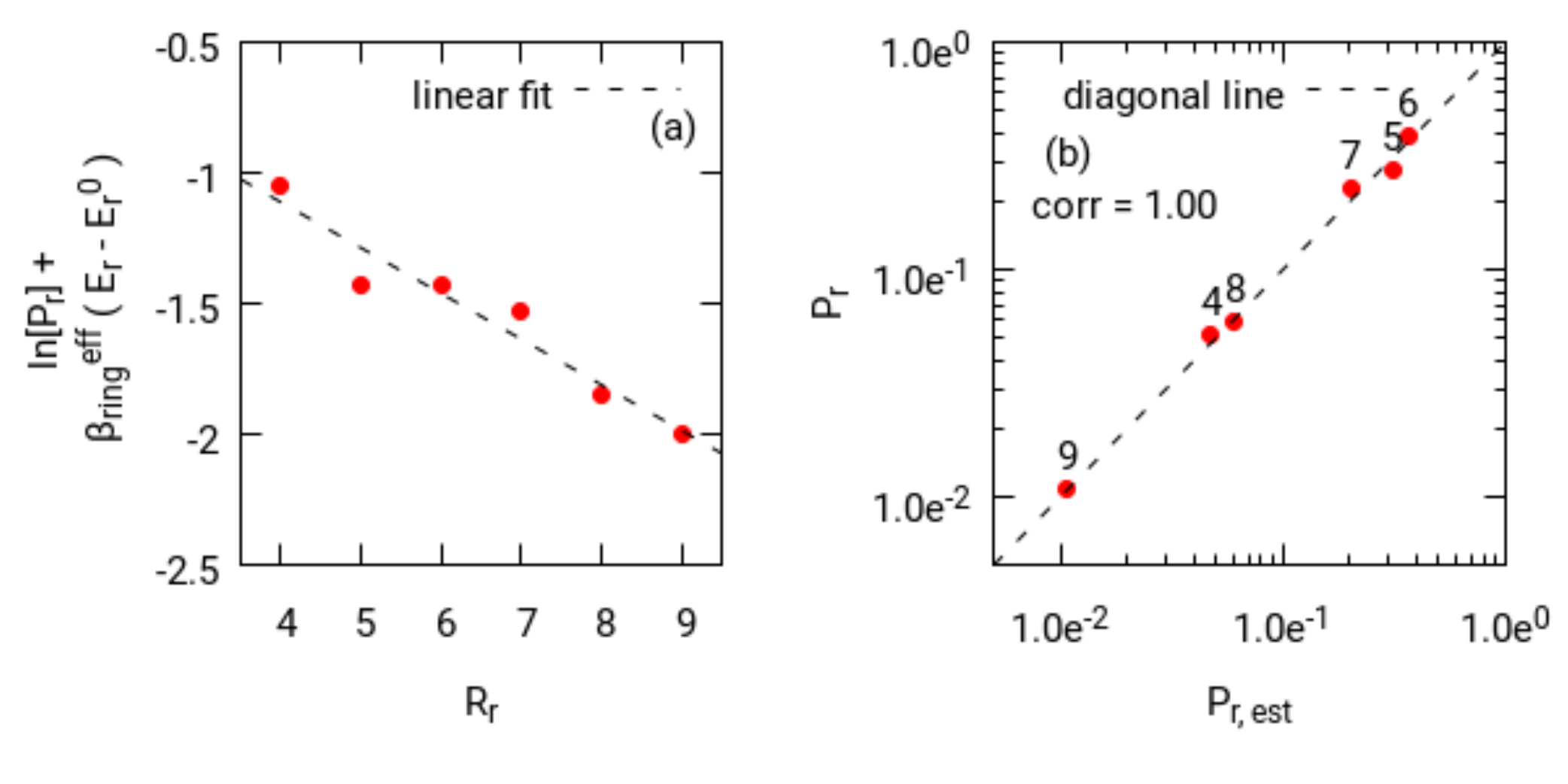}
 \caption{ (a):Size dependence of ring energies with $\beta_{\scriptsize \texttt{ring}}^{\scriptsize \texttt{eff}} = 344.5$, determined from complementary ring analysis \cite{RoyPRL2019}. Since $E_r^0 \propto R_r$ the slope can be directly translated into an estimation of $\eta_{\scriptsize \texttt{ring}}$ = -0.62.  (b): Comparison of the predicted and actual individual ring probabilities. For both graphs, $T = 0.015$ ($\beta = 66.7$).}
 \label{fig:ring_energies}
\end{figure*}

It is possible to derive the $\beta_{\scriptsize \texttt{ring}}^{\scriptsize \texttt{eff}}$ without any interference from $\eta_{\scriptsize \texttt{triplet}}$ if one couples probabilities of ring size $R_r$ and its complementary ring size $R_r^* = 12 - R_r$ together in the Arrhenius plot as

\begin{equation}
 P_r P_r^* \propto e^{-\beta_{\scriptsize \texttt{ring}}^{\scriptsize \texttt{eff}} (E_r + E_r^*)}
 \label{eqn:comp_ring_enegy}
\end{equation}

This procedure was chosen in \cite{RoyPRL2019} so that no additional Lagrange parameter, taking care of the correct average ring size, was necessary. The value of $\beta_{\scriptsize \texttt{ring}}^{\scriptsize \texttt{eff}}$ is approximately 5 times higher than $\beta$. In this work we do not reiterate the physical implications of this difference but rather refer to Ref.\cite{RoyPRL2019}. We just mention that the observation $\beta_{\scriptsize \texttt{ring}}^{\scriptsize \texttt{eff}} \ne 3 \beta$ reflects the presence of non-standard thermodynamics on local scales \cite{RoyPRL2019}.

Now we estimate the complete ring size distribution via  equation (\ref{eqn:ring_new_boltzmann}). We fix the value of $\beta_{\scriptsize \texttt{ring}}^{\scriptsize \texttt{eff}}$ as obtained from the above analysis, based on the complementary rings. The correlation of [$ln(P_r) +  \beta_{\scriptsize \texttt{ring}}^{\scriptsize \texttt{eff}} (E_r - E_r^0)$] with the ring size allows one to obtain the value of $\eta_{\scriptsize \texttt{ring}}$ via linear regression as shown in figure \ref{fig:ring_energies}. We observe a basically perfect correlation. Thus, our general approach equation (\ref{eqn:ring_new_boltzmann}) contains the key elements to express the probabilities in terms of energies and system sizes.

\begin{figure*}[!htb]
\centering
\includegraphics[width=12cm, height=6cm]{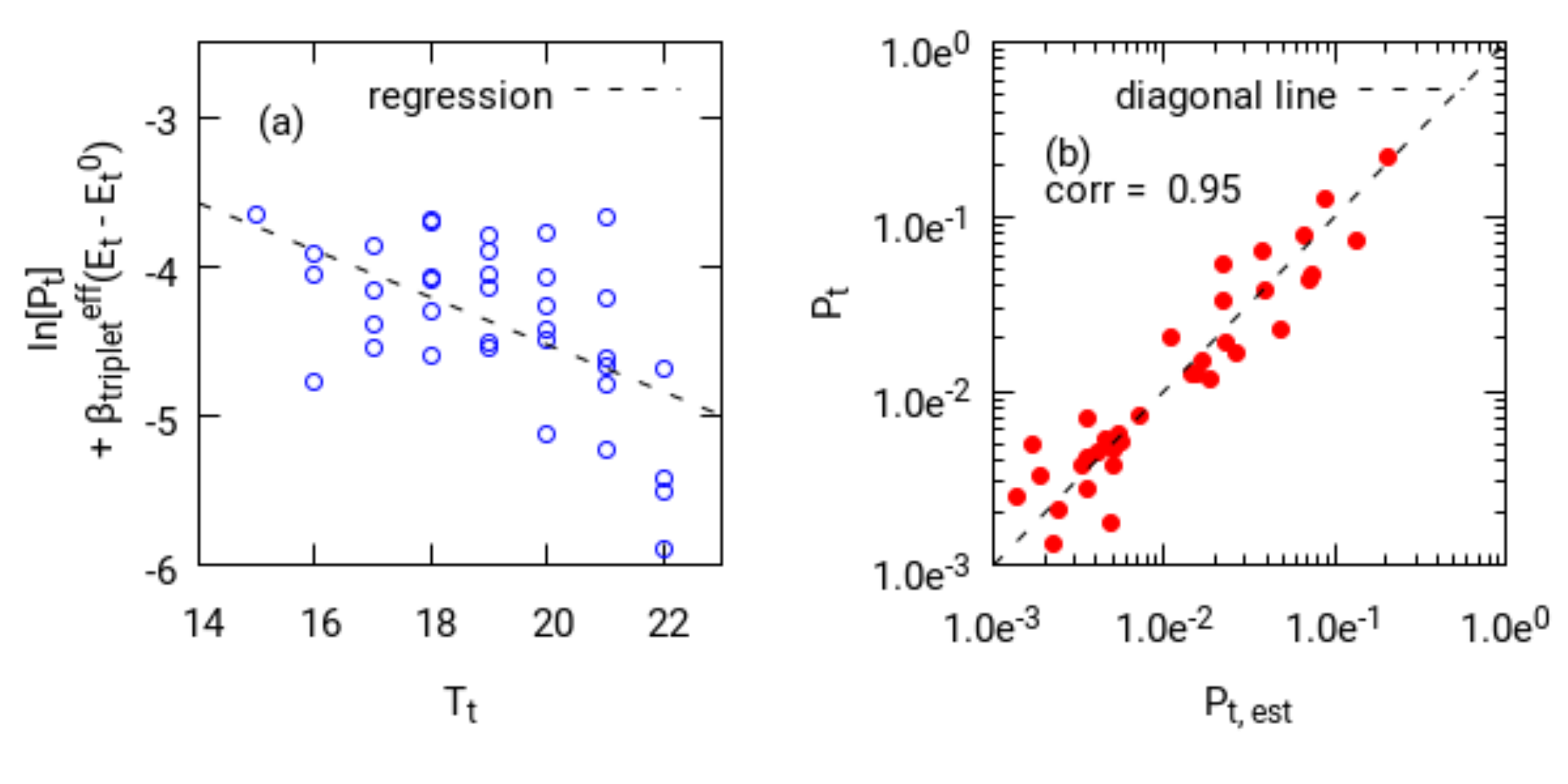}
\caption{(a):Size dependence of triplet energies at $T = 0.015$ with $\beta_{\scriptsize \texttt{triplet}}^{\scriptsize \texttt{eff}} = 182.5$, determined from the complementary triplet analysis \cite{RoyPRL2019}. The regression yields $\eta_{\scriptsize \texttt{triplet}}$ = -0.563 (b): Predicted vs. actual probabilities for individual triplets. All triplets with $P_t > 10^{-3}$ are taken into account.}
\label{fig:triplet_energy}
\end{figure*}

A very similar analysis of complementary sizes can be performed for triplets with $T_t + T_t^* = 36$. The value of $\beta_{\scriptsize \texttt{triplet}}^{\scriptsize \texttt{eff}}$ is about 3 times higher than $\beta$. As discussed in \cite{RoyPRL2019} for triplets this factor is close to what is expected for standard thermodynamics. This reflects the fact that the triplet sizes are naturally much larger than the size of individual rings. Using this value of $\beta_{\scriptsize \texttt{triplet}}^{\scriptsize \texttt{eff}}$ we estimate $\eta_{\scriptsize \texttt{triplet}}$ and use this value to predict the probabilities of the individual triplets; see figure \ref{fig:triplet_energy}. The minor deviations in the prediction of the probabilities may be related to the effects, discussed in the context of equation \ref{eqn:const_average_trip_size}.

\begin{figure*}[!htb]
\centering
 \includegraphics[width=12cm,height=6cm]{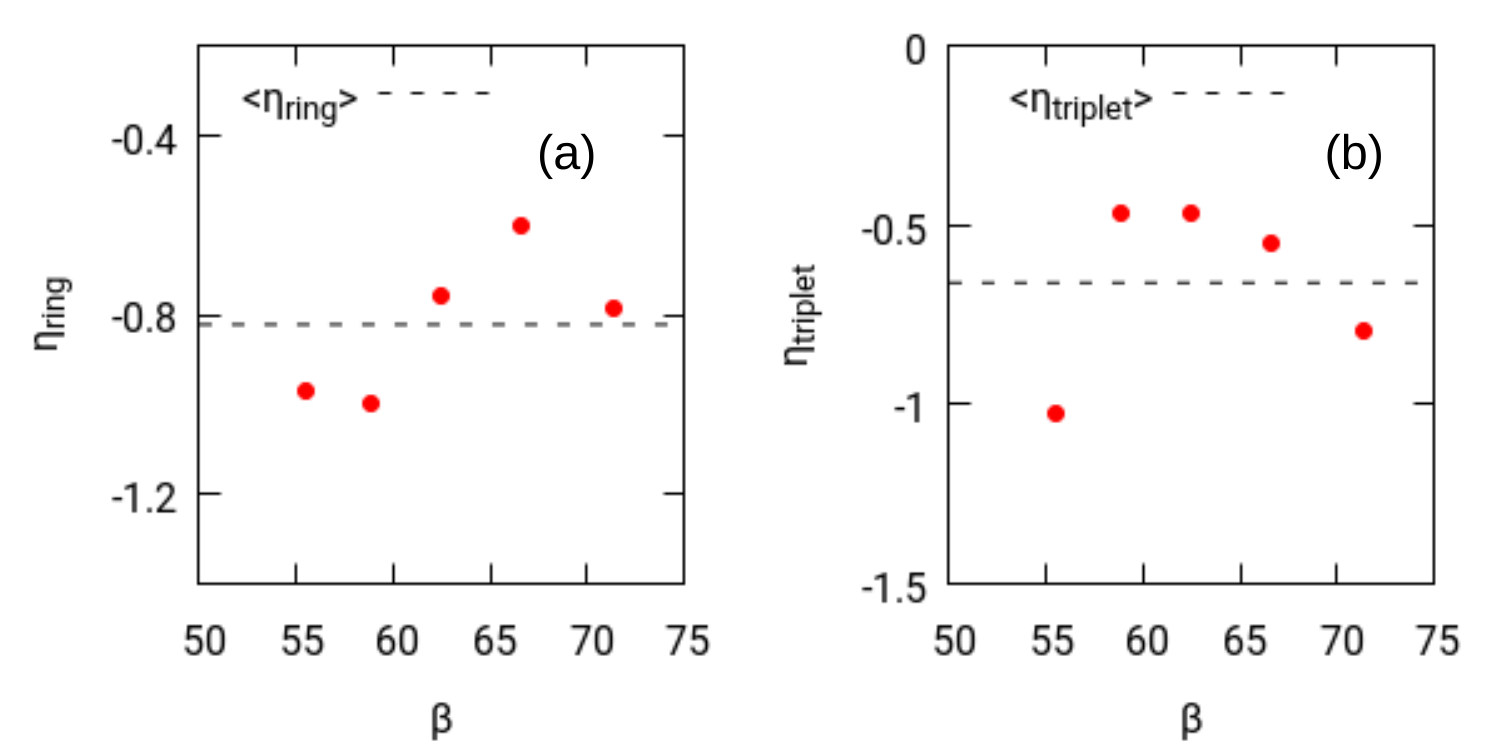}
 \caption{Variation of (a) $\eta_{\scriptsize \texttt{ring}}$ and (b) $\eta_{\scriptsize \texttt{triplet}}$ with inverse temperature ($\beta$). The average value of $<\eta_{\scriptsize \texttt{ring}}>$ and $<\eta_{\scriptsize \texttt{triplet}}>$ reads -0.82 and -0.63, respectively.}
 \label{fig:all_eta}
\end{figure*}

Finally, we repeat this analysis for simulations at different temperatures. The agreement is always very similar. The resulting values of $\eta_{\scriptsize \texttt{ring}}$ and $\eta_{\scriptsize \texttt{triplet}}$ are shown in figure \ref{fig:all_eta}. Despite significant statistical uncertainties, one can conclude that both Lagrange parameters are basically temperature independent. This justifies the specific form, used in  equation (\ref{eqn:ring_new_boltzmann}).

\section{Estimation of energies}
\label{sect:Angle_mismatch_theory}
\subsection{Ring and triplet energies from angle deviations}

So far we used the measured energies to predict the occurrence of different rings and triplets. Here we want to better understand the origin of the ring and triplet energies. In the previous work, the energies have been empirically related to the ideal inner $\angle SiSiSi$ angles of the rings \cite{HeydeJnonCrysSolid2016}. Here we can check and extend this simple approach by using the information from the actual energies and apply this approach to triplets as well. Although, the Yukawa force-field does not contain any explicit angular term, the ring-energies  still depend of the $\angle SiSiSi$, since any angular deformation in a ring will change the distances between the corners. 

Following Rino et al. \cite{VashishtaPhysRevB1993}, we approximate the ring energy parameters with a harmonic function of the angular deviations of the rings. Naturally, the lowest energy state corresponds to a crystal with a six-ring and an inner angle of 120$^\circ$. In general, a symmetric $R_r$-sided ring with equal edge length has an inner angle of $\theta_r = (180^\circ - 360^\circ/R_r)$. Then we estimate the energy as

\begin{equation}
  E_r = R_r a_{\scriptsize \texttt{ring}} (\theta_r - 120^\circ)^2 + R_r \frac{E_6}{6}
  \label{eqn:Angle_mismatch_theory_rings}
\end{equation}

In this representation $a_{\scriptsize \texttt{ring}}$ is an adjustable parameter. The constant guarantees that  equation (\ref{eqn:Angle_mismatch_theory_rings}) exactly holds for $r = 6$. Since $E_r$ results from an average over many different microscopic realizations, we may expect that the impact of the nature of the adjacent rings is small.

\begin{figure*}[!htb]
\centering
\includegraphics[width=8cm, height=8cm]{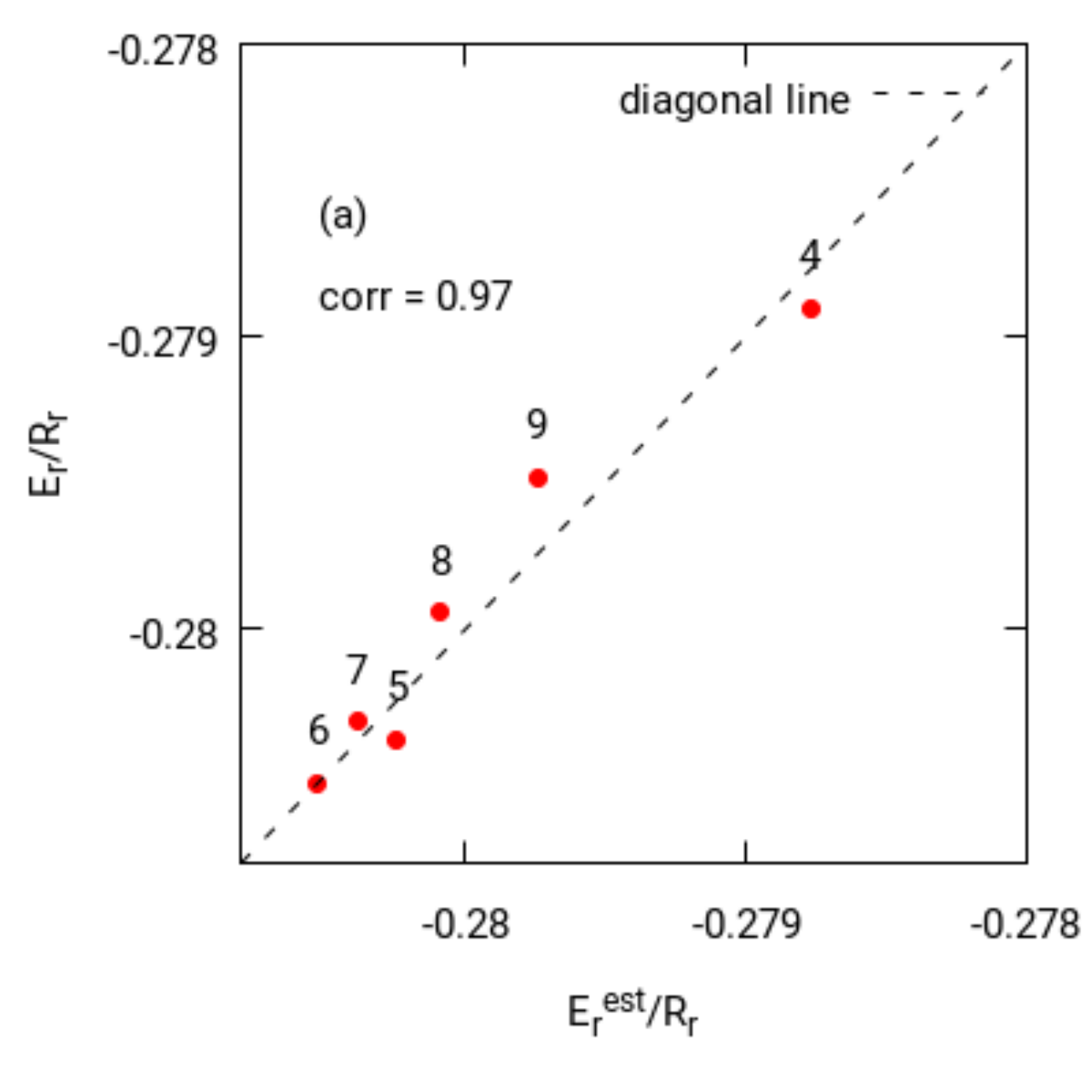}
\caption{Comparison of the energies derived from simulations and equation (\ref{eqn:Angle_mismatch_theory_rings}) at $T = 0.015$. From linear regression we obtain $a_{\scriptsize \texttt{ring}} = 1.95\times 10^{-6}$.}
\label{fig:Angle_mismatch_ring}
\end{figure*}

\begin{figure*}[!htb]
\centering
 \includegraphics[width=12cm,height=6cm]{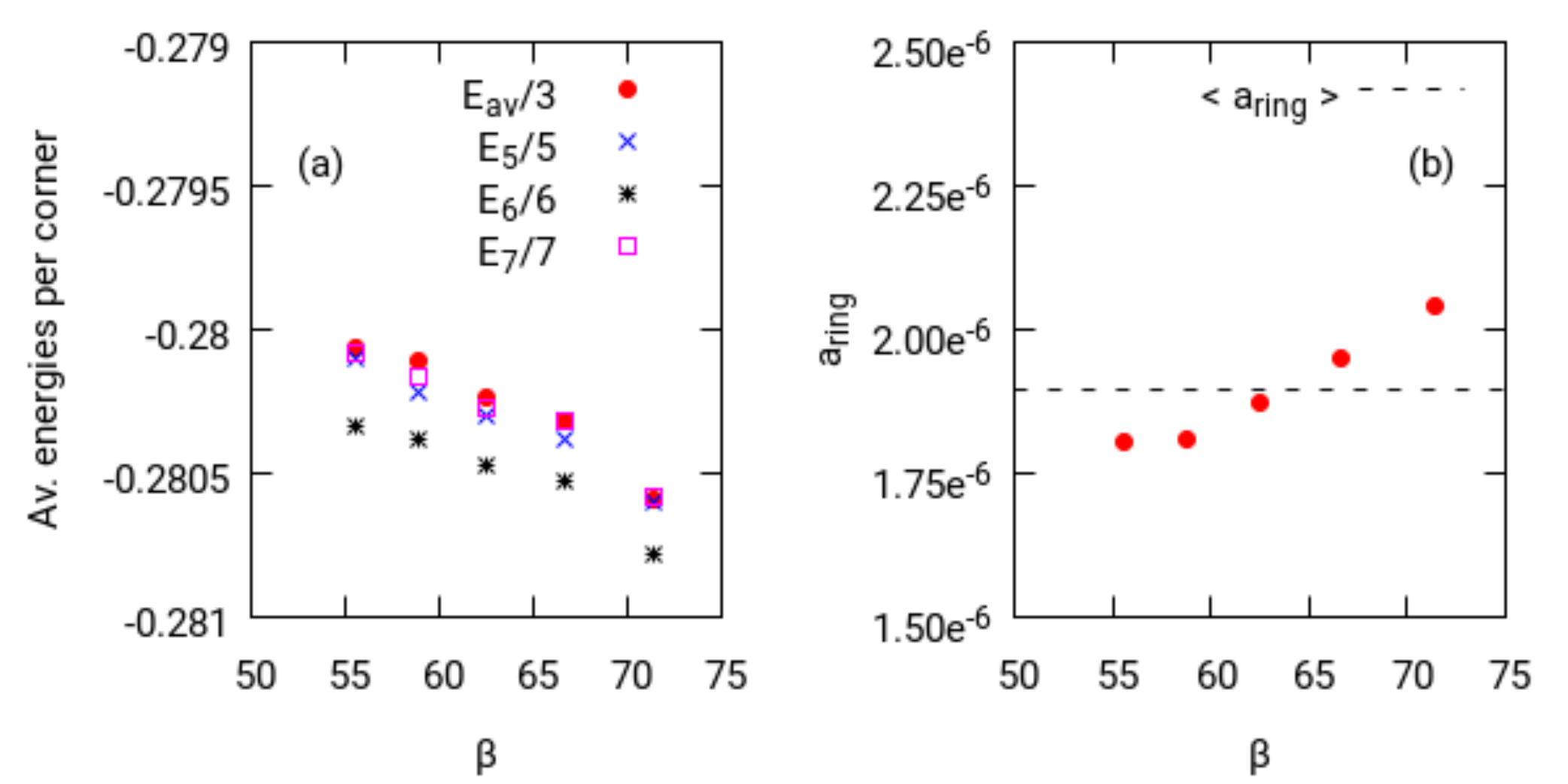}
 \caption{Variation of (a): $E_{\textsf{\scriptsize av}}/3$, $E_5/5$, $E_6/6$, $E_7/7$ and (b): $a_{\scriptsize \texttt{ring}}$ with inverse temperature $\beta$. The average value $<a_{\scriptsize \texttt{ring}}>$ in this temperature range is $1.9 \times 10^{-6}$.}
 \label{fig:a_ring_alltemp}
\end{figure*}

One finds an excellent correlation. Slight deviations are present for $r=4$ and $r=9$. Here the deviations from the ideal angle are strongest and higher-order terms might be required in  equation (\ref{eqn:Angle_mismatch_theory_rings}) to capture the energy in more detail. Naturally, the individual values of $E_6/6$ and $E_r/R_r$ are temperature dependent (see figure \ref{fig:a_ring_alltemp}(a)) since at lower temperatures lower energy states are explored. Furthermore,  the value for $a_{\scriptsize \texttt{ring}}$ (see figure \ref{fig:a_ring_alltemp}(b)) is basically temperature independent, albeit with a residual drift. As clearly seen from  figure \ref{fig:a_ring_alltemp}(a), the temperature dependence of $E_r/R_r$ for $r \ne 6$ can be largely explained by the nearly identical temperature dependence of $E_6/6$.  Thus, the temperature variation of $a_{\scriptsize \texttt{ring}}$ hardly contributes in equation (\ref{eqn:Angle_mismatch_theory_rings}) and thus can be neglected. We would like to stress that the differences between rings, reflected, e.g., by the different occurrence probabilities in the present temperature range \cite{RoyPCCP2018}, is a mere consequence of $a_{\scriptsize \texttt{ring}} > 0$.

Next, we want to predict the energies of the triplets. Three contributions are taken into account. First, we consider the self-correlation of the individual rings.

\begin{equation}
 f_t = \sum_r R_r (\theta_r - 120^\circ)^2
 \label{eqn:Self_correlation_rings}
\end{equation}

\begin{figure*}[!htb]
\centering
\includegraphics[width=4cm, height=4cm]{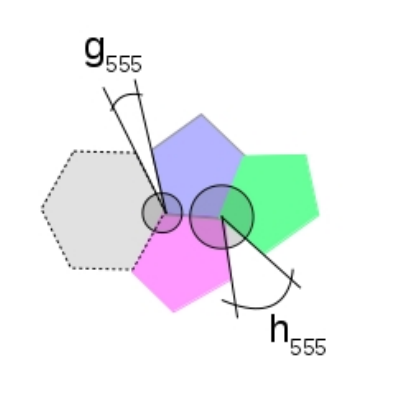}
\caption{A sketch of different angle mismatch functions $h_{555}$ and $g_{555}$ in a hypothetical 555 triplet (red,green,blue). The total angle mismatch from $360^\circ$ at the center is calculated via $h_{555}$. Assuming the average neighbor ring-size of a triplet is six,  the total angle mismatch from $360^\circ$ at the corners next to center is $g_{555}$.}
\label{fig:triplet_angle_error_sketch}
\end{figure*}

Second, we incorporate the extra strain originating from bringing three rings of a triplet together at a common corner. It has been discussed in  \cite{HeydeJnonCrysSolid2016} that the deviation of the total inner ring angles from $360^\circ$; i.e., for a $ijk$ triplet,

\begin{equation}
  h_t = | \theta_i + \theta_j + \theta_k - 360^\circ |
  \label{eqn:Angle_error_triplets}
\end{equation}

is correlated with the actual occurrence of that triplet. Here, we use this term to refine the energy estimation for a triplet. Third, a similar term can be calculated for the immediate neighbors of the central Si particles by assuming an average neighbor ring-size of six (see figure \ref{fig:triplet_angle_error_sketch}) as,

\begin{equation}
 g_t =  | \theta_i + \theta_j  - 240^\circ | +  | \theta_i + \theta_k - 240^\circ | +  | \theta_j + \theta_k - 240^\circ |
 \label{eqn:Angle_error_doublets}
\end{equation}

Specifically, we choose

\begin{equation}
 E_t =   < a_{\scriptsize \texttt{ring}} > f_t + \Delta_t + T_t \frac{E_{666}}{18}
 \label{eqn:Angle_mismatch_theory_triplets}
\end{equation}

where

\begin{equation}
 \Delta_t = b_{\scriptsize \texttt{triplet}} h_t + c_{\scriptsize \texttt{triplet}} g_t
 \label{eqn:Total_angle_error}
\end{equation}

The final term guarantees that  equation (\ref{eqn:Angle_mismatch_theory_triplets}) is exact for the 666 triplet.

\begin{figure*}[!htb]
\centering
\includegraphics[width=12cm, height=12cm]{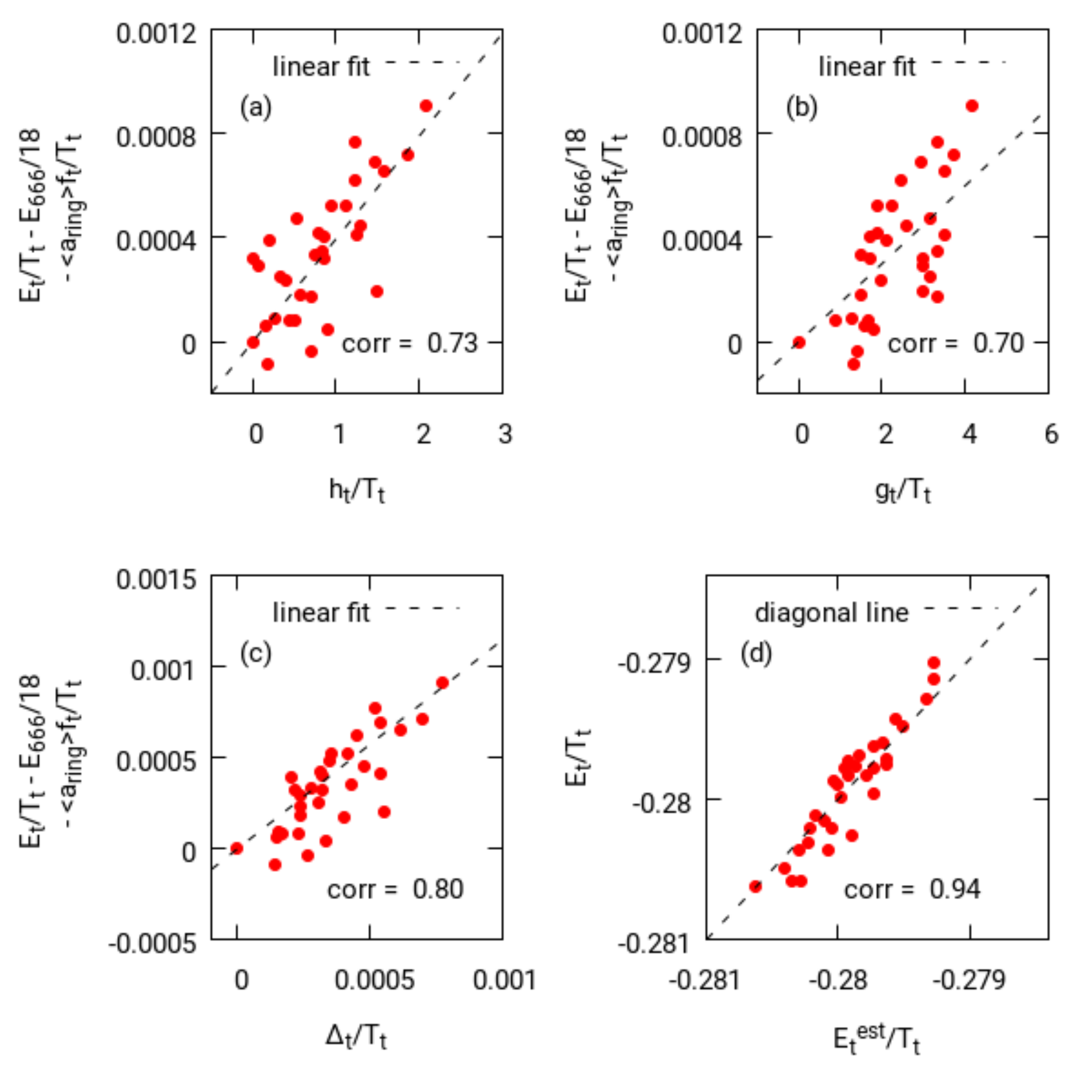}
\caption{Dependence of the residual strain of a triplet with (a): $h_t$, (b): $g_t$, and (c): $\Delta_t$. Included are linear fits. (d):Comparison of the energies derived from simulations and  equation (\ref{eqn:Angle_mismatch_theory_triplets}) at $T = 0.015$. The values of $b_{\scriptsize \texttt{triplet}}$ and $c_{\scriptsize \texttt{triplet}}$, as derived from (c) via regression are $2.25 \times 10^{-4}$ and $7.29 \times 10^{-5}$, respectively. All data are plotted which had $P_t > 2 \times 10^{-3}$ for individual triplets.}
\label{fig:Angle_mismatch_triplet}
\end{figure*}

First, we show in figure \ref{fig:Angle_mismatch_triplet}(a)(b) that both $h_t$ and $g_t$ display a significant correlation with the residual energy $[ E_t - T_t (E_{666} / 18) -<a_{\scriptsize \texttt{ring}}>f_t ] $, just including contributions from the individual rings. Thus, the correlations of the three rings, forming a triplet, have a significant influence on the triplet energy. It also shows that the linear dependence on $h_t$ and $g_t$ is indeed consistent with the data.  Finally, we performed a multiple regression with respect to $b_{\scriptsize \texttt{triplet}}$ and $c_{\scriptsize \texttt{triplet}}$. The result is shown in figure \ref{fig:Angle_mismatch_triplet}(c). The resulting energy estimation is then compared with the actual energy in figure \ref{fig:Angle_mismatch_triplet}(d). Again a high correlation is observed. This proves that  equation (\ref{eqn:Angle_mismatch_theory_triplets}) is indeed a good approximation for estimating the triplet energies. Note that we have used the assumption, see discussion of  equation (\ref{eqn:Angle_error_doublets}), that the triplets are connected to rings of size 6 in their neighborhood. The residual fluctuations indicate the presence of additional inter-triplet effects. The high correlation coefficient of 0.94 directly shows that additional correlations only have a small impact (if at all) on the average energy of a given triplet. We mention in passing (data not shown) that we do not observe any systematic temperature dependence of $b_{\scriptsize \texttt{triplet}}$ and $c_{\scriptsize \texttt{triplet}}$.  Their average values in the considered temperature regime are $b_{\scriptsize \texttt{triplet}} = 2.5 \times 10^{-4}$ and $c_{\scriptsize \texttt{triplet}} = 5.5 \times 10^{-5}$. In analogy to $E_6$ also $E_{666}$ decreases with decreasing temperature. 

\subsection{Estimating the probability distributions}

Next, we study how well knowledge of $a_{\scriptsize \texttt{ring}}$, $b_{\scriptsize \texttt{triplet}}$ and $c_{\scriptsize \texttt{triplet}}$ allows us not only to predict the energies but also the probabilities for the different rings and triplets, respectively. We also use the effective temperature $\beta^{\scriptsize \texttt{eff}}_{\scriptsize \texttt{triplet}}$ from  \cite{RoyPRL2019} and the values of $\eta_{\scriptsize \texttt{triplet}}$ as determined in section \ref{sect:Ring_energy_results}.

\begin{figure*}[!htb]
 \centering
\includegraphics[width=12cm, height=6cm]{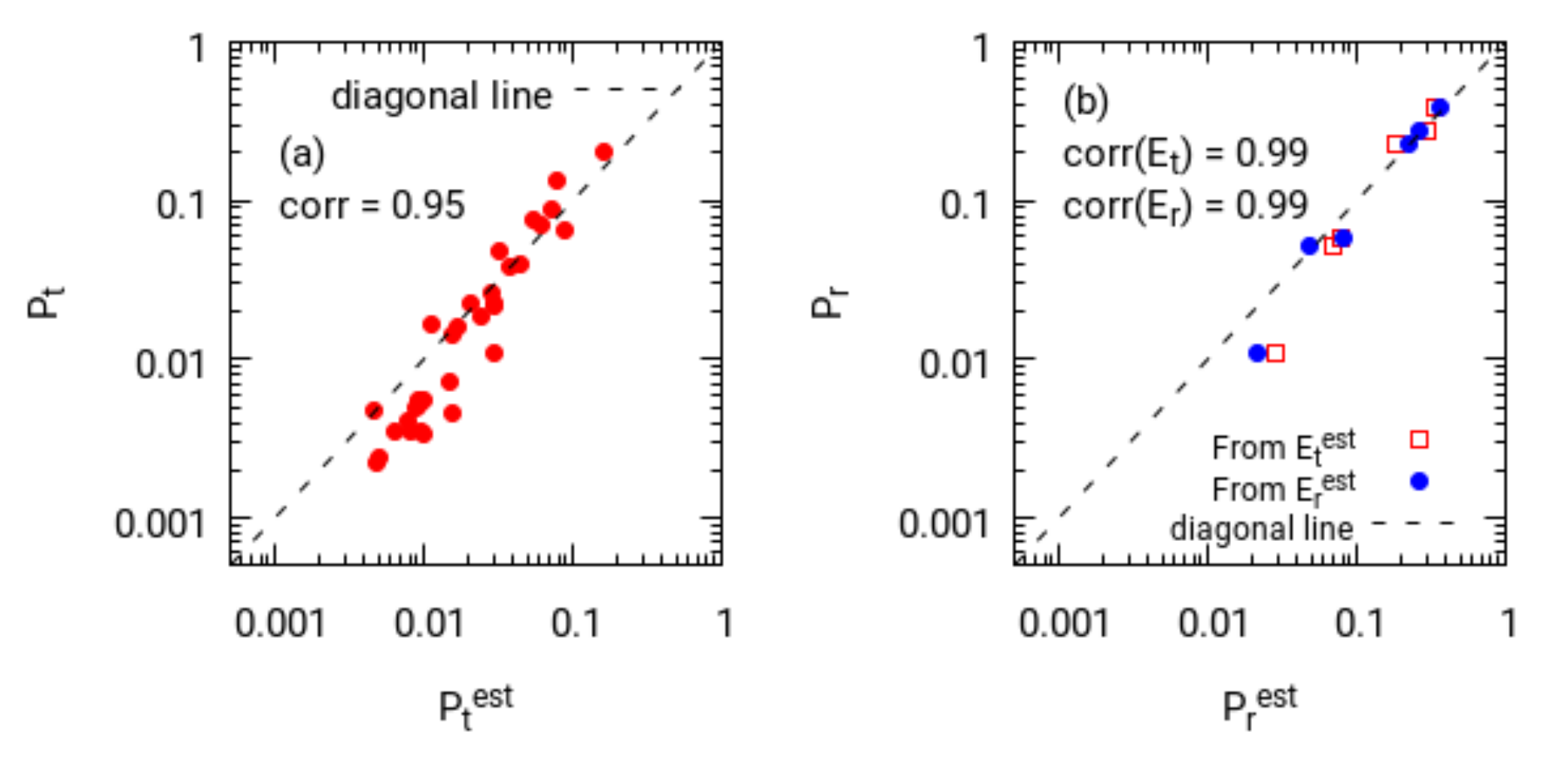}
\caption{Comparison of the estimated and actual (a): triplet probabilities and (b): ring probabilities, at $T = 0.015$. For (a), all data points with $P_t > 0.002$ for both the estimated and the simulated probabilities are plotted. In (b), we show the estimation of ring probabilities from the estimated triplet probabilities (equation (\ref{eqn:ring_prob_empirical})) in empty red squaresin blue dots. The estimation of ring probabilities without the triplet probabilities, i.e. solely from the ring energies (equation (\ref{eqn:Angle_mismatch_theory_rings}) and equation (\ref{eqn:ring_new_boltzmann})), is shown in blue dots.}
\label{fig:P_empirical}
\end{figure*}

We assume that the constants $a_{\scriptsize \texttt{ring}}$, $b_{\scriptsize \texttt{triplet}}$, $c_{\scriptsize \texttt{triplet}}$ and $\eta_{\scriptsize \texttt{triplet}}$ are temperature independent and use their average values for the subsequent calculations. Now, the estimated triplet probability distribution, $P_t^{est}$, can be derived from equation (\ref{eqn:trip_new_boltzmann}). The empirical probabilities for ring distribution ($P_r^{est}$) can be directly derived from $P_t^{est}$ as

\begin{equation}
 P_r^{est} = \frac{2 f_r}{R_r}\sum_i \sum_{j \geq i} P_{ijr}^{est}
 \label{eqn:ring_prob_empirical}
\end{equation}

where `$ f_r$' is the permutation factor given by,

\begin{eqnarray}
  f_r & = 3 \condition{for $r=j=k$} \nonumber \\
      & = 2 \condition{for $r=j$ or $r=k$} \nonumber \\
      & = 1 \condition{for $r \neq j$ and $r \neq k$}
 \label{eqn:cr}
\end{eqnarray}

We can also estimate ring probabilities from  equation (\ref{eqn:Angle_mismatch_theory_rings}) and  equation (\ref{eqn:ring_new_boltzmann}), without using the triplet properties. The results are very similar as shown in figure \ref{fig:P_empirical}(b). Basically, this can be regarded as a consistency check of combination of the two relations, discussed so far (structure $\rightarrow$ energy; energy $\rightarrow$ probability). For further results regarding the consistency check, see appendix A).

\section{Discussion and Outlook}

We have shown that it is possible to estimate, on the one hand, the energies of rings and triplets based on the size of the involved rings and, on the other hand, to relate the energies to the actual probabilities for the case of 2D-silica. Since we are dealing with simulated data, we can directly take the actual energies from the simulated configurations.

The Lagrange parameter $\eta$, taking care of the correct ring size is basically temperature independent (both for rings and triplets). As already indicated above, the formulation of the two constraints ( equation (\ref{eqn:ring_new_boltzmann}) and  equation (\ref{eqn:trip_new_boltzmann})) is not unique. For example, we might have taken the two constraints $\sum_r N_r E_r = E_{\textsf{\scriptsize tot}}$ and $\sum_r N_r E_r^0 = E_{\textsf{\scriptsize tot}}$. Naturally, the prediction of the probabilities would be identical. However, in this case, both Lagrange parameters would be strongly temperature dependent.

Furthermore, we would like to stress that in agreement with the suggestions in literature, the ring/triplet energies are strongly related to the angle deviation of the average inner angles of the rings from a 6-ring. The key focus of the present work is a clear-cut numerical identification of these effects for 2D-silica. Again, we find that the prefactors, obtained from a straightforward regression procedure, are basically temperature independent. This is equivalent to the observation that the average energies of the rings of different sizes have a nearly identical temperature dependence. Naturally, based on these parameters it is possible to express parameters such as the Aboav-Wearie parameter, describing the properties of random networks (see appendix A). Thus, the present identification of appropriate system-specific adjustable parameters, describing the energies as well as the Lagrange parameters to predict the probabilities, may serve as an underlying description of the Aboav-Wearie parameter for 2D-silica. Naturally, for different systems, other parameters might emerge.

The successful prediction of ring and triplet energies, just based on the size of the involved rings and their direct correlation, may suggest that this mapping also holds for different networks where energies are not readily available. It may also be possible to extend this method to generate large correlated networks via straightforward Monte-Carlo simulations. In the future, we hope to estimate the ring-distribution in 3D bulk-silica with a similar methodology. It will also be interesting to see how the 3D-silica rings are dependent on the internal angle-mismatch, analogous to the case of 2D-silica.

\ack{There are no conflicts of interests to declare. We thank the NRW Graduate School of Chemistry and the University of M\"unster for providing the necessary funding. We thank Markus Heyde, Fritz-Haber-Institute of the Max-Planck-Society, Berlin for important discussions on the concept of triplets and suggestions about the manuscript.}

\section*{References}

\bibliographystyle{iopart-num}
\providecommand{\noopsort}[1]{}\providecommand{\singleletter}[1]{#1}%
\providecommand{\newblock}{}

\appendix

\section{Consistency check of the estimated energies from the angular mismatch calculations}

We  further check to which degree the estimated energies from equation \ref{eqn:Angle_mismatch_theory_triplets} are consistent with the observed triplet probabilities. Two important parameters related to the intra-ring correlations are chosen, the correlation factor (denoted `effective' probability in \cite{RoyPCCP2018}) and the Aboav-Wearie parameter \cite{AboavMetallography1983}.

The correlation factor of a $t = ijk$ triplet, $P^{\scriptsize \texttt{eff}}_t$,  can be defined as,

\begin{equation}
 P_t^{\scriptsize \texttt{eff}} = \frac{P_t}{P_i P_j P_k}
 \label{eqn:effective_prob}
\end{equation}

\begin{figure*}[!htb]
\centering
 \includegraphics[width=8cm, height=8cm]{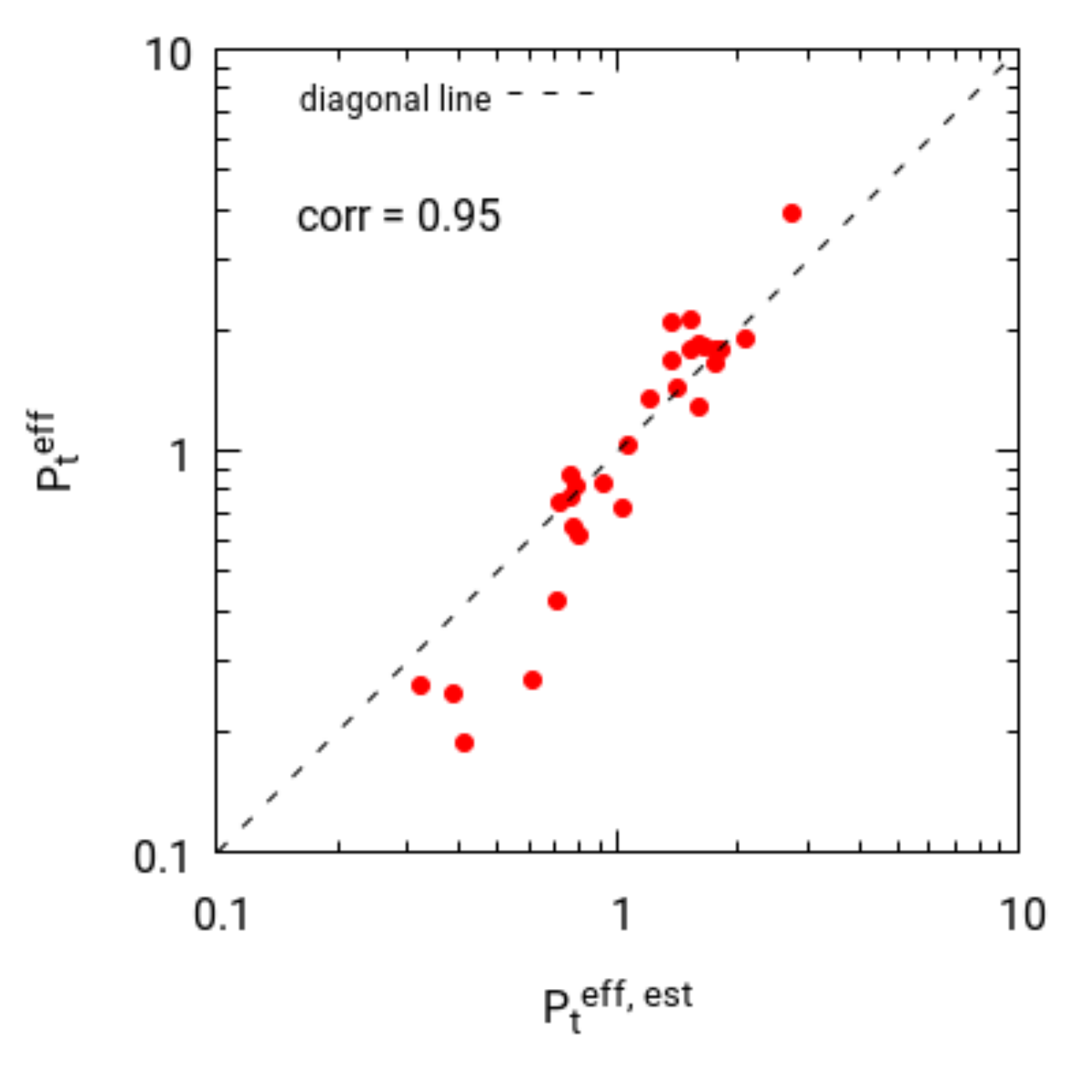}
 \caption{Comparison of effective triplet probabilities between estimated angle-mismatch theory and simulated results, at $T = 0.015$. All data points with both $P_t > 0.002$ and $P_t^{predicted} > 0.002$ are plotted for both estimated and simulated system. The estimated predicted probabilities are calculated via  equation (\ref{eqn:ring_prob_empirical}).}
 \label{fig:P_eff_empirical}
\end{figure*}

$P_t^{\scriptsize \texttt{eff}}$ shows how sensitive the occurrence of rings are with their neighbors as compared to an uncorrelated system with the same ring probabilities. Thus, $P_t^{\scriptsize \texttt{eff}}$ gives us a measure of the inherent stability of the triplet. Here, the empirical ring-probabilities are derived from the triplet distribution itself. Again, an excellent correlation is observed in figure \ref{fig:P_eff_empirical}.

\begin{figure*}[!htb]
\centering
\includegraphics[width=8cm, height=8cm]{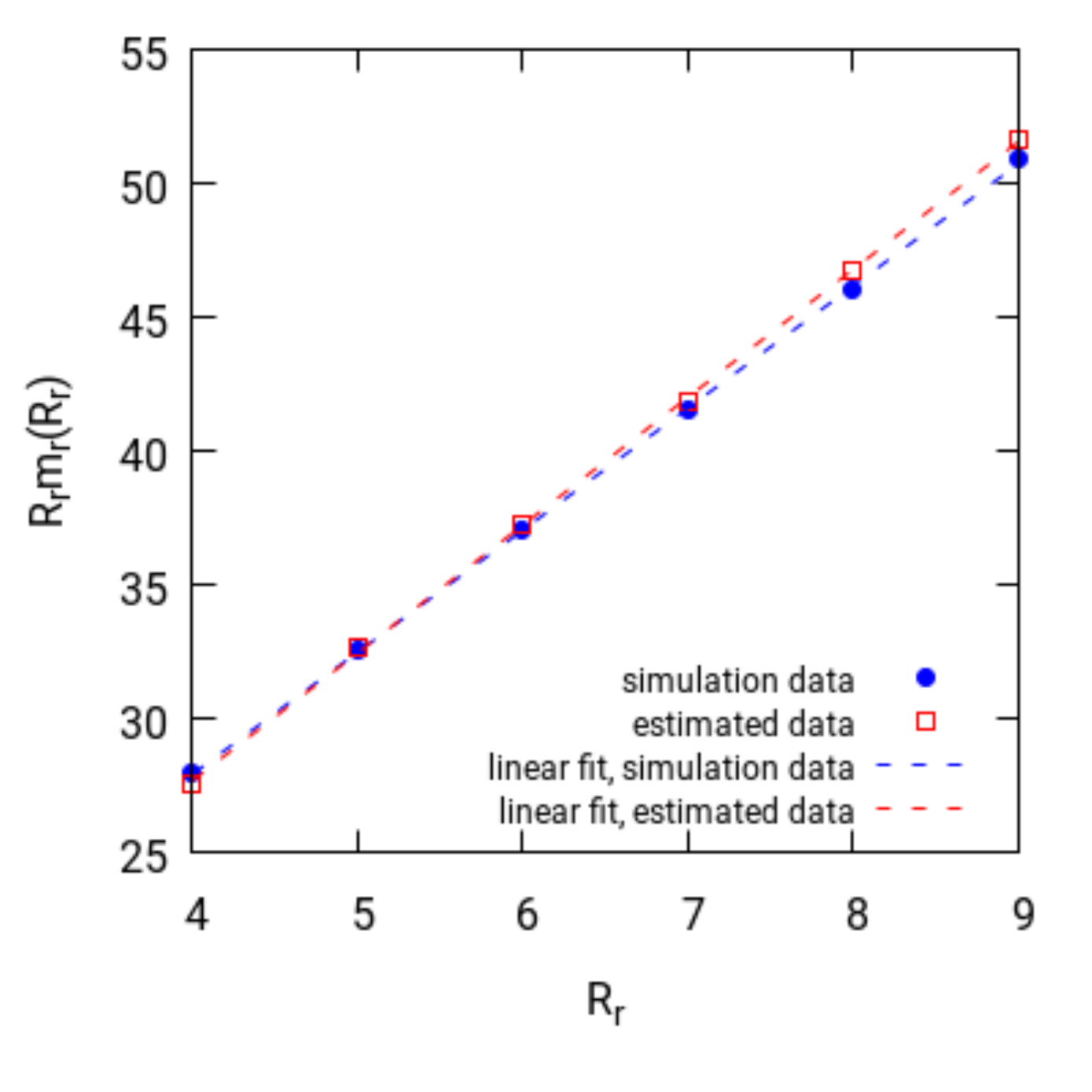}
\caption{Comparison of Aboav-Wearie plot for simulation data and estimated data at $T = 0.015$. The Aboav-Wearie parameter, calculated from a linear fit is 1.43 for the simulated data and 1.22 for the data, estimated by the angle-mismatch approach.}
\label{fig:Aboav_Wearie_comparison}
\end{figure*}

We have also calculated the Aboav-Wearie parameter from figure \ref{fig:Aboav_Wearie_comparison} for both simulated data and estimated data. To calculate the average neighbor ring size ( $m_r$ ) of a ring-size ($R_r$), we calculate the estimated probability of finding two neighbor ring-sizes, i and j ($j  \geq i$), in any position of space ($P_{ij}^{est}$) as ,

\begin{equation}
 P_{ij}^{est} = \frac{1}{3}\sum_r f_r P_{ijr}^{est}
 \label{eqn:duplet_prob_empirical}
\end{equation}

Following this, we can calculate

\begin{equation}
 m_r(R_r) = \frac{\sum_i R_iP_{ir}^{est} + R_rP_{rr}^{est}}
                 {\sum_i P_{ir}^{est} + P_{rr}^{est}}
 \label{eqn:average_ring_neighbor_size}
\end{equation}

We use the estimated probabilities from figure \ref{fig:P_empirical} to calculate $P_{ij}^{est}$ in equation \ref{eqn:duplet_prob_empirical}. We found that the Aboav-Wearie parameter is closely reproduced by the angle-mismatch theory. The success of theory in deriving Aboav-Wearie parameter proves that this parameter indeed has a strong relationship with the inner ring angles. However, derivation of a more direct relationship of Aboav-Wearie parameter and angles is found to be more complex and thus not included in this work.

\section{Triplet energies}

\begin{center}

\tablefirsthead{
 \hline
 Triplets & Size       & Probability  & Av. Energy     & Standard error of the mean \\
              & ($T_t$)   & ($P_t$)        & ($E_t / T_t$)          & ($SE_t$/$10^{-5}$) \\
  \hline}
 \tablehead{
 \hline
  \multicolumn{5}{|l|}{Table continued from previous page} \\
 \hline
 Triplets & Size       & Probability  & Av. Energy     & Standard error of the mean \\
              & ($T_t$)   & ($P_t$)        & ($E_t / T_t$)          & ($SE_t$/$10^{-5}$) \\
  \hline}
\tabletail{
 \hline
 \multicolumn{5}{|l|}{Table continued to next page} \\
 \hline}
 \tablelasttail{}\bottomcaption{}

 \topcaption{Average energies and probabilities of various triplets at T=0.015.}
 \label{tbl:Ring_av_energyt}
\begin{supertabular}{|c|c|c|c|c|}
  \hline
446 & 14 &   0.24*$10^{-5}$    & -0.27605   &   225.139  \\
447 & 15 &   6.10*$10^{-5}$    & -0.27667   &   60.536   \\
448 & 16 &   7.90*$10^{-5}$    & -0.27742   &   29.908   \\
449 & 17 &   3.91*$10^{-5}$    & -0.27755   &   70.741   \\
455 & 14 &   2.85*$10^{-5}$    & -0.27677   &   66.107   \\
456 & 15 &   1.02*$10^{-3}$    & -0.27847   &   15.005   \\
457 & 16 &   3.55*$10^{-3}$    & -0.27914   &   8.377    \\
458 & 17 &   4.03*$10^{-3}$    & -0.27929   &   8.812    \\
459 & 18 &   2.26*$10^{-3}$    & -0.27902   &   13.517   \\
466 & 16 &   5.53*$10^{-3}$    & -0.27948   &   6.711    \\
467 & 17 &   0.023                       & -0.27975   &   3.879    \\
468 & 18 &   0.015                       & -0.27971   &   5.408    \\
469 & 19 &   5.07*$10^{-3}$    & -0.27943   &   9.043    \\
477 & 18 &   0.017                       & -0.27996   &   5.510    \\
478 & 19 &   0.016                       & -0.27978   &   5.909    \\
479 & 20 &   5.54*$10^{-3}$    & -0.27960   &   10.858   \\
488 & 20 &   3.53*$10^{-3}$    & -0.27962   &   12.986   \\
489 & 21 &   1.88*$10^{-3}$    & -0.27956   &   17.813   \\
499 & 22 &   0.14*$10^{-3}$    & -0.27889   &   59.573   \\
555 & 15 &   0.64*$10^{-3}$    & -0.27938   &   10.768   \\
556 & 16 &   0.022                       & -0.28026   &   3.546    \\
557 & 17 &   0.038                       & -0.28036   &   3.028    \\
558 & 18 &   0.022                       & -0.28020   &   4.526    \\
559 & 19 &   4.59*$10^{-3}$    & -0.27973   &   9.961    \\
566 & 17 &   0.088                       & -0.28058   &   1.958    \\
567 & 18 &   0.202                       & -0.28058   &   1.491    \\
568 & 19 &   0.066                       & -0.28031     &   2.894    \\
569 & 20 &   0.011                       & -0.27998   &   8.177    \\
577 & 19 &   0.074                       & -0.28036     &   2.917    \\
578 & 20 &   0.039                       & -0.28016   &   4.245    \\
579 & 21 &   7.30*$10^{-3}$    & -0.27977   &   10.475   \\
588 & 21 &   4.99*$10^{-3}$    & -0.27983   &   13.135   \\
589 & 22 &   1.70*$10^{-3}$    & -0.27973   &   22.017   \\
599 & 23 &   0.10*$10^{-3}$    & -0.27928   &   52.302   \\
666 & 18 &   0.072                       & -0.28062   &   2.120    \\
667 & 19 &   0.132                       & -0.28049   &   2.043    \\
668 & 20 &   0.026                       & -0.28012   &   5.192    \\
669 & 21 &   3.37*$10^{-3}$    & -0.27978   &   16.371   \\
677 & 20 &   0.049                       & -0.28020   &   4.063    \\
678 & 21 &   0.019                       & -0.27990   &   6.533    \\
679 & 22 &   3.56*$10^{-3}$    & -0.27982   &   14.663   \\
688 & 22 &   1.38*$10^{-3}$    & -0.27973   &   21.499   \\
689 & 23 &   0.60*$10^{-5}$    & -0.27958     &   33.258   \\
699 & 24 &   1.80*$10^{-5}$    & -0.27926   &   124.215  \\
777 & 21 &   4.83*$10^{-3}$    & -0.27987   &   13.962   \\
778 & 22 &   2.41*$10^{-3}$    & -0.27969   &   17.568   \\
779 & 23 &   0.69*$10^{-3}$    & -0.27965     &   39.525   \\
788 & 23 &   0.62*$10^{-3}$    & -0.27977   &   44.291   \\
789 & 24 &   0.21*$10^{-3}$    & -0.28021   &   50.819   \\
799 & 25 &   0.63*$10^{-5}$    & -0.27920   &   247.662  \\
888 & 24 &   7.94*$10^{-5}$    & -0.27983   &   60.01    \\
889 & 25 &   0.67*$10^{-5}$    & -0.27998   &   57.587   \\
899 & 26 &   0.12*$10^{-5}$    & -0.27970   &   0.102    \\
\hline
\end{supertabular}
\end{center}

\end{document}